\documentclass{eas}
\usepackage{graphicx}
\begin{document}

\def\HI{H{\sc i} }
\def\HII{H{\sc ii} }
\def\OIII{O[{\sc iii}] }
\def\Msun{M$_{\odot}$ }
\def\OH{{12\-+log(O/H)} }
\def\lNO{log(N/O) }
\def\cd{chemo-dynamical }
\FirstPage{113}

\title{Massive stars: their contribution \\
to energy and element budget \\
in chemo-dynamical galaxy evolution} 
\runningtitle{Hensler: Galactic energy and element budget from massive stars}
\author{Gerhard Hensler}
\address{Institute of Astronomy, University of Vienna, 
T\"urkenschanzstr. 17, A--1180 Vienna, Austria,
\email{hensler@astro.univie.ac.at}}

\begin{abstract}

Here results of numerical radiation hydrodynamical simulations are
presented which explore the energetic impact of massive stars on 
the interstellar medium. 
We study the evolution of the ambient gas around isolated massive 
stars in the mass range between 15 and 85 \Msun in order to analyze the 
formation of structures and further the transfer and deposit of the 
stellar wind and radiation energy into the circumstellar medium until 
the stars explode as a supernovae. 
The derived energy transfer efficiencies are much smaller than
analytically estimated and should be inserted into chemo-dynamical
evolutionary models of galaxies as appropriate parameter values. 
As an additional issue the element release in the Wolf-Rayet phases 
and its detectability have been investigated for comparison with observations.

\end{abstract}
            \maketitle
\section{Introduction}

Besides external effects structure and evolution of galaxies is primarily
determined by the energy budget of its interstellar medium (ISM). Reasonably,
also the ISM structure and both its radiative and kinetic energy contents
are determined by the energy deposit of different sources, on the one hand, 
and by the energy loss by means of radiative cooling on the other.
Mainly massive
stars contribute significantly to this structure formation like e.g.
cavities, holes, and chimneys in the \HI gas and superbubbles of hot gas
(Recchi \& Hensler \cite{rec06a}).
On large scales the energy release by massive stars triggers the matter
circulations via galactic outflows from a gaseous disk and galactic winds
(Recchi \& Hensler, this volume).
By this, also the chemical evolution is affected thru the loss of 
metal-enriched gas from a galaxy (see e.g. Recchi et al. \cite{rec06}). 

Supernova (SN) explosions as an immediate consequence of star formation (SF) 
stir up the ISM by the expansion of hot bubbles, deposit turbulent energy 
into the ISM and, thereby, regulate the SF again (Hensler \& Rieschick 
\cite{hens02}). 
This negative energy feedback is enhanced at low gravitation because 
the SN energy exceeds easily the galactic binding energy and drives 
a galactic wind. Vice versa, SN and stellar wind-driven bubbles sweep up 
surrounding gas and can, by this, excite SF self-propagation as a 
positive feedback mechanism (e.g. Ehlerova et al. \cite{ehl97}, 
Fukuda \& Hanawa \cite{fuk00}). 

Since the numerical treatment of the galactic \cd evolution, although 
when it deals with two gas phases, cannot resolve the ISM spatially 
sufficiently, for a detailed consideration of small-scale processes
the chemo-dynamical modeling has to apply parametrizations of 
plasmaphysical processes. It should, however, be repeatedly emphasized 
that an appropriate \cd description cannot vary those parameters arbitrarily, 
because they rely on results from theoretical, numerical, and/or 
empirical studies and are by this strongly constrained. 
This has been described and applied to different levels of \cd models 
(for an overview see Hensler, \cite{hens03} and references therein, 
and more recently Hensler et al., \cite{hens04}; Harfst et al., \cite{har06}).

As an example, analytical consideration of heating and cooling balance 
by K\"oppen et al. (\cite{koe95}) making use of the cooling by 
collisionally excited radiation (B\"ohringer \& Hensler \cite{bh89}) 
yields that the SF rate under reasonable ISM states is 
dependent on the square of the gas density with a slight influence of
the heating efficiency. This fact is surprising: although the ionizing 
radiative fluxes of massive stars are pretty well known from static 
stellar atmosphere models (see e.g. Thompson \cite{thom84}) or more
sophisticated wind atmospheres (Kudritzki \& Puls \cite{kud00}), the
fraction of the radiative energy that is thermalizing the gas is uncertain. 

In addition, massive stars act also dynamically on their surrounding
ISM by a strong stellar wind (Kudritzki \& Puls \cite{kud00}). 
Although the stellar wind power $L_w = \frac{1}{2}\dot{M}_w v_w^2$ 
with the mass-loss rate ($\dot{M}_w$) and the terminal wind velocity ($v_w$) 
can be easily evaluated from model and observational values,  
the fraction of the wind luminosity which is transferred e.g.\ into 
thermal energy, the so-called thermal energy transfer efficiency, or
into turbulent energy is not obvious from principles.

\section{Theory of radiative and stellar wind bubbles}

Within the framework of the simplest theoretical approach an O star
suddenly ``turns on'' in a constant density medium at rest and begins to
ionize its surroundings. The reaction of the medium to the stellar photons
is well-known and has been described in detail in standard textbooks.
With the particular knowledge of the time evolution of the ionization front 
at distance 
from a star of age $\tau$ and with 
respect to the initial Str\"omgren radius $R_s$ which depends on the 
Lyman continuum photon luminosity, 
the \HI number density of the ambient medium,
and the isothermal sound speed of the ionized gas 
 Lasker (\cite{las67}) derived the kinetic energy of the expanding 
swept-up shell 
$E_{\mathrm{k}}$,
   the ionization energy stored in the \HII region 
$E_{\mathrm{i}}$,
and the thermal energy of warm gas in the \HII region
$E_{\mathrm{t}}$
(see also Freyer et al. \cite{frey03}, hereafter: Paper I).
As an appropriate approximation the temperature in the \HII region
needed in the set of formulae
can be set to 8000 K.

Also for stellar wind bubbles (SWBs) the radial structure can be
analytically derived (see Weaver et al. \cite{weav77} and Paper I) 
and, furthermore, also the kinetic energy as
$   E_{\mathrm{k}}
      = \frac{3}{11}\,L_{\mathrm{w}}\,\tau $
and the thermal energy of the hot gas as
$ E_{\mathrm{t}}
      = \frac{5}{11}\,L_{\mathrm{w}}\,\tau $.
Besides heat conduction and cooling in the hot bubble here
also neglected is the fact that part of the thermal energy might be
used for collisional ionization of gas. Thus, the results for
$E_{\mathrm{k}}$ and $E_{\mathrm{t}}$ are upper limits.

Because these analytical derivations cannot account for aspherical
structure formation and turbulence within the SWB which both consume
stellar energy and alter the energy fractions numerical models are
needed to investigate the massive stellar energy deposit into the
ISM. For a comprehensive tabulation and comparison of former models
the interested reader is referred to Paper I.

\section{The model}

The numerical code applied for this purpose and used to obtain the 
results presented here is described in Paper I. 
The hydrodynamical equations are solved together with the transfer of 
H-ionizing photons on a 2d cylindrical grid. 
The time-dependent ionization and recombination of hydrogen is calculated 
each timestep, and we carefully take stock of all the important energy 
exchange processes in the system. A comprehensive description of the 
algorithm is given in Yorke \& Kaisig (\cite{yor95}), 
Yorke \& Welz (\cite{yor96}), and in Paper I.

This code is applied to the evolution of massive stars with 4 different masses: 
15 \Msun (Kroeger, Freyer, Hensler \& Yorke, in prep. as Paper IV of this series), 
35 \Msun (Freyer et al. \cite{frey06}: Paper II of this series),
60 \Msun (Paper I), and
85 \Msun (Kroeger et al. \cite{kroe06b}: Paper III of this series). 
Because we started the project also at first for comparison with the 60 \Msun 
model published by Garcia-Segura et al. (\cite{garcia96a}) we adopted basically 
the same stellar parameters applied by them from stellar evolutionary models 
and from observations. 
As initial conditions of the ambient medium we also use the same undisturbed 
background gas layer as Garcia-Segura et al. 
with \HI number density $n_0 = 20\, cm^{-3}$ and temperature $T_0$ = 200 K.
This yields a thermal pressure ``typical'' for the ISM at the solar 
galacto-centric distance. 
Also the 35 \Msun model was aimed for comparison with Garcia-Segura et al. 
(\cite{garcia96b}) and used their stellar parameters along its life. 
Because from this group no evolutionary tracks and stellar parameters
of other stellar masses became available yet, however, for the 15 \Msun and 
the 85 \Msun models we were compelled to apply stellar evolutionary models 
from Schaller et al. (\cite{schal92}).
Although this makes the sample inhomogeneous, the extensive computational time 
necessary for the models only allowed until now comparison of the two
60 \Msun star model sources. As a first result, we obtained for the 
Schaller et al. model a by one order of magnitude weaker wind luminosity 
(Hensler, Kroeger, \& Freyer, in prep.). 
Furthermore, the evolutionary sequence of Wolf-Rayet (WR) and Luminous Blue Variable 
(LBV) phases are reverse from one group to the other in the sense that
the WR phase precedes the LBV phase in the Langer models.

In our models the stellar parameters $\dot{M}_w$, $v_w$, effective temperature 
($T_{\mathrm{eff}}$), and photon luminosity in the Lyman continuum 
($L_{\mathrm{LyC}}$) are 
time-dependent boundary conditions which drive and govern the evolution of 
the circumstellar gas.

We start our calculations with the sudden turn-on of the zero-age main-sequence 
stellar radiation field and stellar wind. For the surrounding gas we neglect
any photo-dissociation of molecular content here which would in reality precede
the photo-ionization stage (e.g. Kroeger, Hensler, \& Freyer, in prep. as 
Paper V of this series).

  \section{Results}

    \begin{figure}
    \label{fig1}
    \includegraphics[width=7cm,angle=-90]{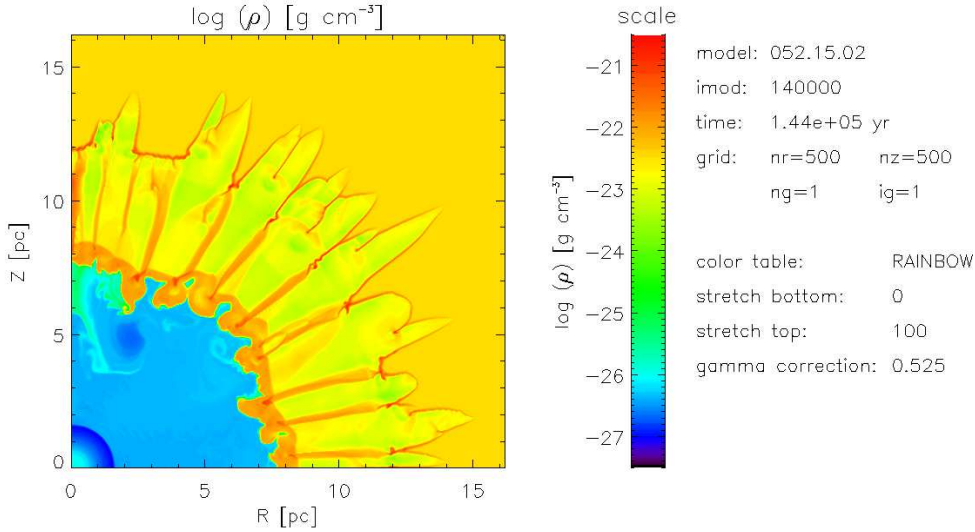} 
    \caption{Density plot of the wind-blow and radiation-driven bubble 
    around a 60 \Msun main-sequence star at an age of 140\,000 years. } 
    \end{figure}

As a first result, the winds structure the surrounding gas by dynamical 
shock front instabilities during its vehement expansion. These lead to
the formation of condensations within the wind shell which, on the other hand, 
have higher optical depth for the ionizing photons. Finger-like extentions of 
the \HII region surrounding the SWB are most characteristicly formed for the
60 \Msun model (see Fig.1) from photons penetrating deeper into the ambient
ISM. In addition, photo-evaporation peels off the inner boundary of the 
wind shell and drives mass loading and turbulence.

   \begin{figure}[h]
    \label{fig2}
    \includegraphics[width=7cm,angle=-90]{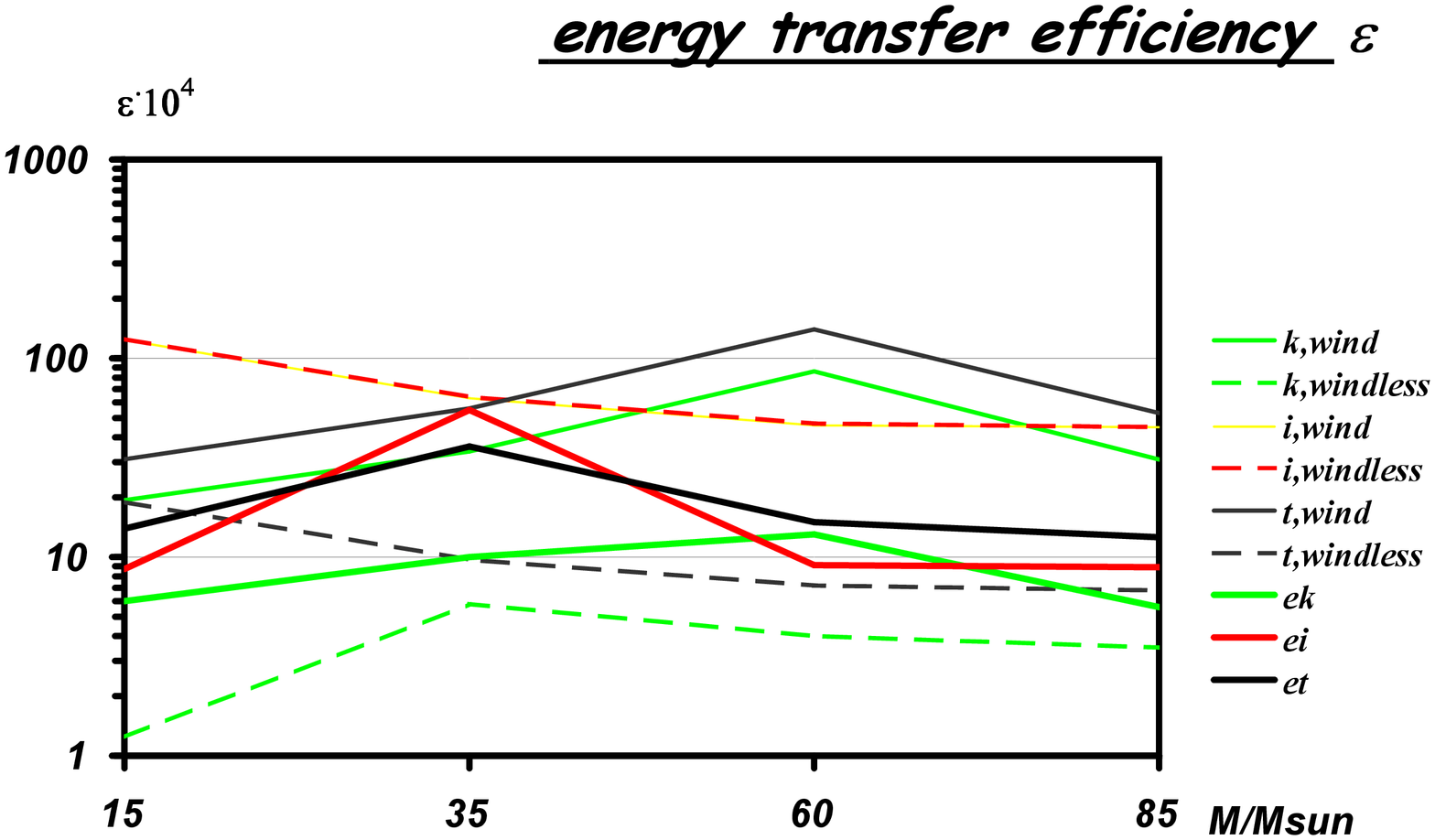} 
    \caption{Energy transfer efficiencies $\varepsilon_{\mathrm{k}},  
    \varepsilon_{\mathrm{i}}$, and $\varepsilon_{\mathrm{t}}$ for kinetic,
    ionizing photon, and thermal energies for the 15, 35, 60, and 85 \Msun models.
k,windless and k,wind as well as for i and t, respectively, represent the analytical
values according to equ.\ref{eq_eps_hii} and its extention by $L_{\mathrm{w}}$,
while $ek, ei, et$ give the efficiency values at the end of the stellar lifetimes. } 
  \end{figure}

Another relevant question addressedto these models is their energy deposit
into the ISM. Here we can only briefly mention a few major issues.
Over the whole evolution one can trace the energy released in different forms
as kinetic $E_{\mathrm{k}}$, photo-ionizing $E_{\mathrm{i}}$, and thermal
  $E_{\mathrm{t}}$ energies, respectively, by book keeping so that at every 
stellar age $\tau$ each integrated energy form released can be compared with 
that inherently available in the computational domain. The ratio of deposited
stellar energy to the increase in one form compared to the initial state
can be defined as an {\it energy transfer efficiency} $\varepsilon$.
As the simplest approach, $\varepsilon$ for pure Lyman continuum radiation 
into the other energy forms is then descibed by the analytical approach of
  \begin{equation}
    \varepsilon = \frac{E}{\tau\,\langle L_{\mathrm{LyC}} \rangle }\ ,
    \label{eq_eps_hii}
  \end{equation}
where $E$ can be any of $E_{\mathrm{k}}$, $E_{\mathrm{i}}$, and
$E_{\mathrm{t}}$, respectively.
For radiation plus wind energy this nominator has to be extended 
accordingly to
( $\langle L_{\mathrm{LyC}} \rangle + \langle L_{\mathrm{w}} \rangle $). 

These $\varepsilon$ values have been derived for all four massive star models
and their energy forms (Hensler, Freyer, \& Kroeger, 2007, in prep. 
as Paper VI of this series) 
and are exhibited in Fig.2. Not surprisingly, the windless values are 
below the wind models except the ionization efficiency $\varepsilon_{\mathrm{i}}$.
Nevertheless, the average $\varepsilon$ only exceed one percent for the thermal 
efficiency of the 60 \Msun wind model. As already mentioned, there the
structure formation is strongest and might be reduced applying 
the Schaller et al. (\cite{schal92}) model. 

The unexpected decrease of $\varepsilon_{\mathrm{i}}$ (the upper dashed
curve in Fig.2) with increasing stellar mass is caused by the growing 
strength of stellar winds with mass that have
the dynamical effect to compress the ambient shells much stronger so that
the recombination is enhanced. This means, ionization energy is more
efficiently lost for larger stellar masses. 

Comparing the windless i.e. pure photo-ionizing models with analytical
results e.g. by Lasker (\cite{las67}) of about one percent, 
our efficiencies fall short by more than one order of magnitude even
to below 0.1 percent (Fig.2). $\varepsilon_{\mathrm{t}}$ reaches more 
than one percent only for 60 \Msun. 
As another important issue, it should be mentioned that the final 
efficiencies ($ek, ei, et$ in Fig.2) are located significantly below the 
values integrated over the stellar life because of the waeker wind at
the last stages. 

The efficiencies derived from our models are much less than expected 
and of fundamental importance for the energy budget of galaxies during
their evolution. They should be easily inserted into chemo-dynamical
evolutionary codes because integrating the stellar energy release by 
massive stars over their concedingly short lifetimes without taking detailed
heating vs. cooling processes into account would falsify the treatment
of the ISM significantly.
A comprehensive discussion is in preparation as Paper VI.

Finally, a byproduct of these studies should be mentioned, namely, that the
elements release in the WR phase of most massive stars, in particular of carbon, 
although at first incorporated into the hot SWB gas are enabled to cool 
by mixing effects and become observable as warm 
ionized \HII gas (Kroeger et al. \cite{kroe06a}).

\vspace{-0.3cm}
 \acknowledgements
  The author thanks Danica Kroeger and Tim Freyer for their strong
engagement to this project and the organizers for this gorgeous 
and most exciting conference. 
  \endacknowledgements


\end{document}